\documentclass[10pt]{iopart}
\usepackage{newtxtext,newtxmath}
\usepackage{subfigure}
\usepackage{graphicx}
\usepackage{xcolor}

\usepackage[
     backend=biber,
     style=nature,
     natbib=true,
     url=false, 
     doi=false,
     eprint=false
  ]{biblatex}
 
\begin{document}\maketitle
\title{Sub-Doppler laser cooling and magnetic trapping of natural-abundance fermionic potassium.}

\author{Mateusz Bocheński and Mariusz Semczuk}

\address{Institute of Experimental Physics, University of Warsaw, Pasteura 5, 02-093 Warsaw, Poland}
\ead{msemczuk@fuw.edu.pl}
%\vspace{10pt}
%\begin{indented}
%\item[]February 2023
%\end{indented}

\begin{abstract}
We demonstrate the largest number of $^{40}$K atoms that has ever been cooled to deeply sub-Doppler temperatures in a single-chamber apparatus without using an enriched source of potassium. With gray molasses cooling on the $D_1$-line following a standard $D_2$-line magneto-optical trap, we obtain $3\times10^5$ atoms at 10(2)~\textmu K. We reach densities high enough to measure the temperature via absorption imaging using the time-of-flight method. We magnetically trap a mixture of $m_F=-3/2,-5/2$ and $-7/2$ Zeeman states of the $F=7/2$ hyperfine ground state confining $5\times10^4$ atoms with a lifetime of 0.6~s or $\sim$$10^3$ atoms with a lifetime of 2.8~s -- depending on whether the temperature of the potassium dispensers was chosen to maximize the atom number or the lifetime. The background pressure-limited lifetime of 0.6~s is a reasonable starting point for proof-of-principle experiments with atoms and/or molecules in optical tweezers as well as for sympathetic cooling with another species if transport to a secondary chamber is implemented.

Our results show that unenriched potassium can be used to optimize experimental setups containing $^{40}$K in the initial stages of their construction, which can effectively extend the lifetime of enriched sources. Moreover, the demonstration of sub-Doppler cooling and magnetic trapping of a relatively small number of potassium atoms might influence experiments with laser-cooled radioactive isotopes of potassium. 
\end{abstract}

%\submitto{\jpb}
%\maketitle
\ioptwocol

\section{Introduction} 

Observation of the first degenerate Fermi gas in 1999~\cite{1}, only four years after the successful creation of a Bose-Einstein condensate~\cite{2}, expanded the scope of experimental platforms where phenomena are predominantly governed by quantum statistics~\cite{3}. The choice of fermions for laser cooling is rather limited if compared to the number of available bosonic species. Among alkali atoms, only lithium and potassium have long-lived isotopes obeying Fermi statistics, $^6$Li and $^{40}$K. Laser cooling of the latter benefits from the availability of high-power CW lasers, nowadays primarily based on telecom technology~\cite{4} or tapered amplifiers~\cite{5}. However, working with $^{40}$K has a major drawback: natural potassium contains only 0.012\% of this isotope. Pioneering works that demonstrated laser cooling and trapping of several thousands $^{40}$K atoms used single-chamber setups (like in this work) with a natural source of potassium~\cite{6, 7}. These early results made it abundantly clear that an isotopically enriched source would be needed to provide a sufficiently large number of atoms to implement evaporative cooling. Since then, almost all experiments using fermionic potassium have relied on sources enriched to 3\%--6\%~\cite{8,9}. The main issue with enriched potassium is its high price and limited availability -- since the late 1990s, the price has gone up by more than an order of magnitude. As a result, experiments with enriched fermionic potassium do not use Zeeman slowers and almost exclusively rely on 2D MOTs~\cite{10} or double-stage MOTs~\cite{1} as a source of pre-cooled atoms. As opposed to other alkali species, single-chamber apparatuses are rarely used with $^{40}$K, even though a single-chamber design with a source located near the trapping region has enabled studies of the BEC-BCS crossover regime with $^6$Li~\cite{11}. In recent years, some efforts have been made to bypass the need for potassium enrichment. Both a Zeeman slower~\cite{12} and a 2D MOT~\cite{13} have been demonstrated, but they have not been widely adopted by the community. 

In this work we demonstrate the largest number of $^{40}$K atoms that has ever been cooled to deeply sub-Doppler temperatures in a single-chamber apparatus without using an enriched source of potassium. We achieve state-of-the-art final temperatures and phase-space densities, comparable with experiments that use 2D MOTs~\cite{10} or double-stage MOTs~\cite{1} as a source of pre-cooled atoms. Even though the final number of atoms is rather small, we are near the limit of what can be obtained in a single-chamber apparatus using a natural abundance potassium. As a proof-of-principle measurement, we trap 5$\times10^4$ atoms in a magnetic trap, sufficiently many to enable sympathetic cooling by another species~\cite{12} if the magnetically trapped clouds are transferred to a second chamber with a much better vacuum. Our results also show that during the construction of a new experimental setup for experiments with $^{40}$K, cheap natural potassium sources can be used for optimization, thus effectively prolonging the useful lifetime of the enriched source.

\section{Experimental setup} 

The experimental chamber is based on a fused silica glass cell without any antireflection coatings. Ultra-high vacuum is maintained by a 55~l/s ion pump (VacIon Plus 55 Noble Diode) supported by a titanium sublimation pump (fired no more than twice per year). The vacuum chamber is designed for laser cooling of cesium and potassium atoms; therefore, we use two dispensers per species as atomic sources ( SAES Getters Cs/NF/5.5/17 FT10 for Cs and K/NF/3.1/17 FT10 for K). We use potassium with a natural abundance of isotopes, that is, the content of $^{40}$K is only $\sim$0.012$\%$. All dispensers are located about 7~cm from the center of the magneto-optical trap (MOT). We use an LED emitting at the central wavelength of 370~nm (FWHM $\sim$10~nm) to increase the MOT loading rate due to the light-induced atomic desorption (LIAD) process~\cite{14}. This method of increasing the additional free-atom load in the trapping region allows us to release less atoms from the dispensers into the setup, thus minimizing the deterioration of the vacuum quality. The base pressure (i.e., when the dispensers have not been turned on for several days) reaches $10^{-11}$~mbar.

The quadrupole field for the magneto-optical and the magnetic trap is created by a pair of coils in a nearly anti-Helmholtz configuration. The coils are arranged such that the axis of the strongest gradient $B'_{\mathrm{axial}}$ is in the horizontal plane. Throughout the paper, we report the radial gradient $B'_{\mathrm{r}}$ because it sets the effective trapping potential in the vertical direction.

The laser system is designed to maximize the available power by limiting double passes through acousto-optic modulators (AOMs) after the final amplification stage. In general, the design allows us to switch between potassium isotopes ($^{39}$K, $^{40}$K or $^{41}$K) on a sub-millisecond time scale and to implement sub-Doppler cooling on the $D_1$ line for all isotopes. Here, we discuss only the core elements of the laser system relevant to the current work on $^{40}$K. The details of the entire laser system will be the subject of further publication.
\begin{figure}[ht]
\centering

\includegraphics[width=0.9\columnwidth]{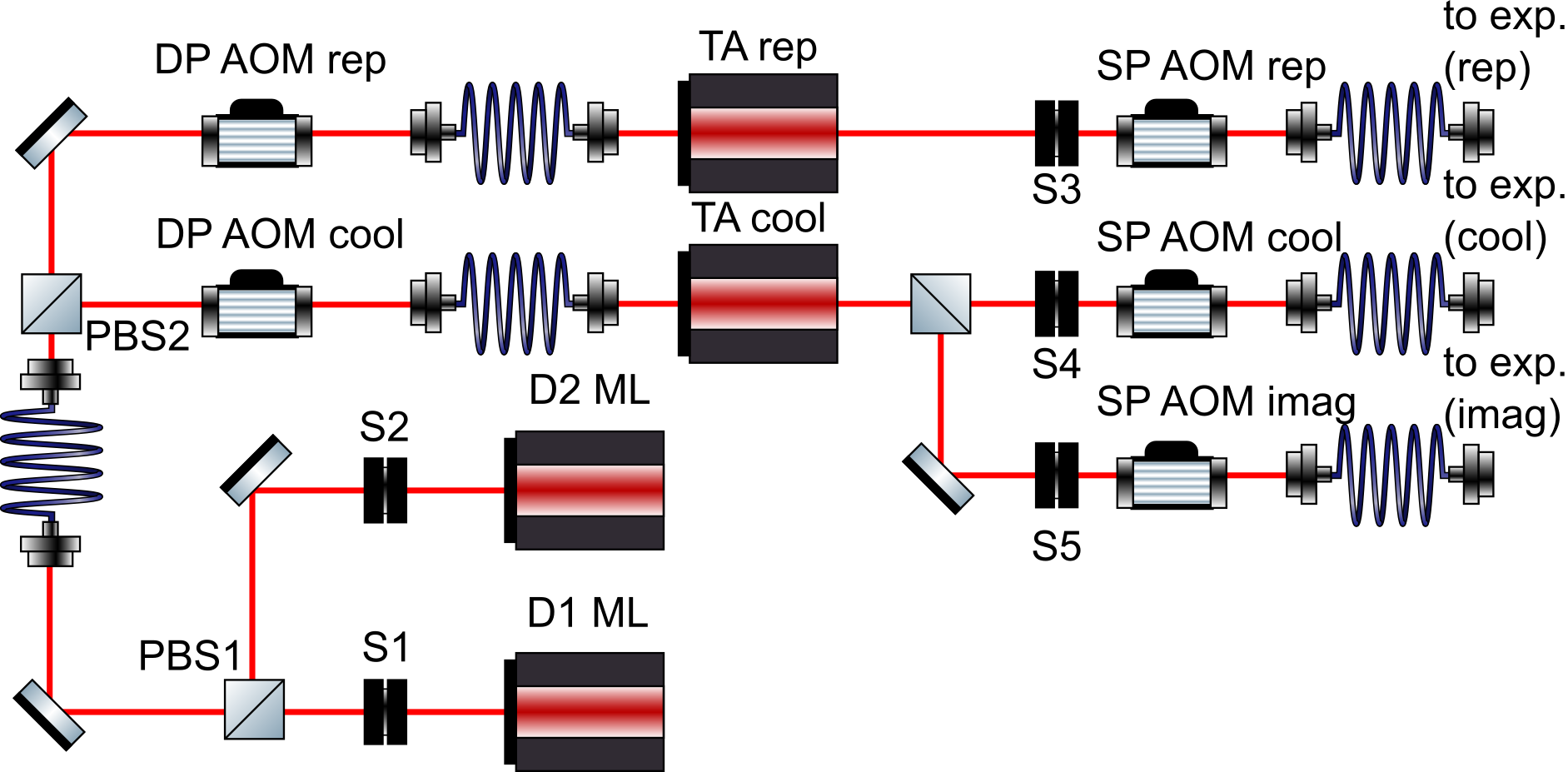}
\caption{Simplified scheme of the $^{40}$K laser cooling system. D1~ML and D2~ML: master lasers (Toptica TA pro) stabilized to crossover transitions of the $D_1$ and the $D_2$ line in $^{39}$K, respectively; DP~AOM$_{\mathrm{cool}}$ and DP~AOM$_{\mathrm{rep}}$: double pass acousto-optic modulators providing total frequency shifts $+(700$--$850)$~MHz and $-(700$--$800$~MHz), respectively,  TA$_{\mathrm{cool}}$ and TA$_{\mathrm{rep}}$: tapered amplifiers (Toptica), SP~AOM$_{\mathrm{cool}}$ and SP~AOM$_{\mathrm{rep}}$: single pass acousto-optic modulators operating at fixed frequencies of $+80$~MHz and $-80$~MHz, respectively, which work as fast optical shutters, S1--S5: mechanical shutters.} 
\label{fig:laser_system_simplified_K}
\end{figure}

Figure~\ref{fig:laser_system_simplified_K} shows a simplified layout of the laser system. We use two master oscillator tapered amplifiers (Toptica TA pro) stabilized to crossover transitions in $^{39}$K, one on the $D_1$ line and the other on the $D_2$ line. The beams from both master lasers are combined on a polarizing beam splitting cube (PBS1) and are coupled into the same single mode, polarization maintaining fiber with perpendicular polarizations. The PBS2 splits each of the $D_1$ and $D_2$ beams into two paths, called cooling and repumping paths, where two double pass acousto-optic modulators DPAOM$_{\textrm{cool}}$ and DPAOM$_{\textrm{rep}}$ provide frequency tuning around optimized values of $+350$~MHz and $-370$~MHz, respectively, before seeding dedicated tapered amplifiers TA$_{\textrm{cool}}$ and TA$_{\textrm{rep}}$ (both are Toptica TA pro models, but we only use tapered amplifiers from these systems). Mechanical shutters S1 and S2 (Uniblitz LS6Z2) are used to choose whether the $D_1$ or $D_2$ line light is amplified. Single-pass acousto-optic modulators SPAOM$_{\textrm{cool}}$ and SPAOM$_{\textrm{rep}}$ provide final frequency shifts of $+80$~MHz and $-80$~MHz, respectively, and work as fast optical shutters. To eliminate light leakage caused by the finite extinction of the AOMs, mechanical shutters S3-S5 (Uniblitz LS6Z2) are used. Our design allows us to reach all frequencies required for efficient cooling on both $^{40}$K lines.
In fermionic potassium, the splitting of the P$_{3/2}$ state is sufficiently large to allow efficient (compared to bosonic potassium isotopes) cooling and compression using the $D_2$ line, therefore we do not need to implement a $D_1$\&$D_2$-line compressed MOT as used in experiments with $^{39}$K and $^{41}$K~\cite{15, 16}. The design of our laser system provides an additional feature, namely the phase coherence of the cooling and the repumping beams, which might be useful for driving Raman transitions between hyperfine states. Phase coherence has been shown to increase the cooling efficiency on the $D_2$ line for some species~\cite{17}, but it is not clear whether it provides any benefit in our setting since the role of phase coherence for gray molasses on the $D_1$ line has not been studied in the literature. 

The single-chamber design of our vacuum system and the use of unenriched dispensers as a source of $^{40}$K heavily restrict the possibility of optimizing the MOT. In particular, we are forced to choose between maximizing the atom number and maximizing the lifetime. These two are strongly coupled unlike in setups where a 2D MOT or a Zeeman slower is used, even if these sources rely on natural abundance potassium~\cite{13,12}.
We use three retro-reflected MOT beams, each collimated with a standard 30~mm diameter lens to a $1/e$ diameter of about 25~mm. This size guarantees that there is no noticeable distortion of the beams caused by the glass cell walls, which are separated by 30~mm. After passing through the glass cell, the diameter of the beams is decreased to $\approx$9~mm with a telescope to fit through $0.5\,''$ wave plates. This choice is made to avoid purchasing custom-made, 30~mm diameter waveplates which are rather costly if they have to be achromatic (our setup is also used for laser cooling of cesium). The retro-reflected beams are slightly focused to partially compensate for reflection losses on the uncoated glass cell walls.
\begin{figure}[ht]
\centering
\includegraphics[width=0.8\columnwidth]{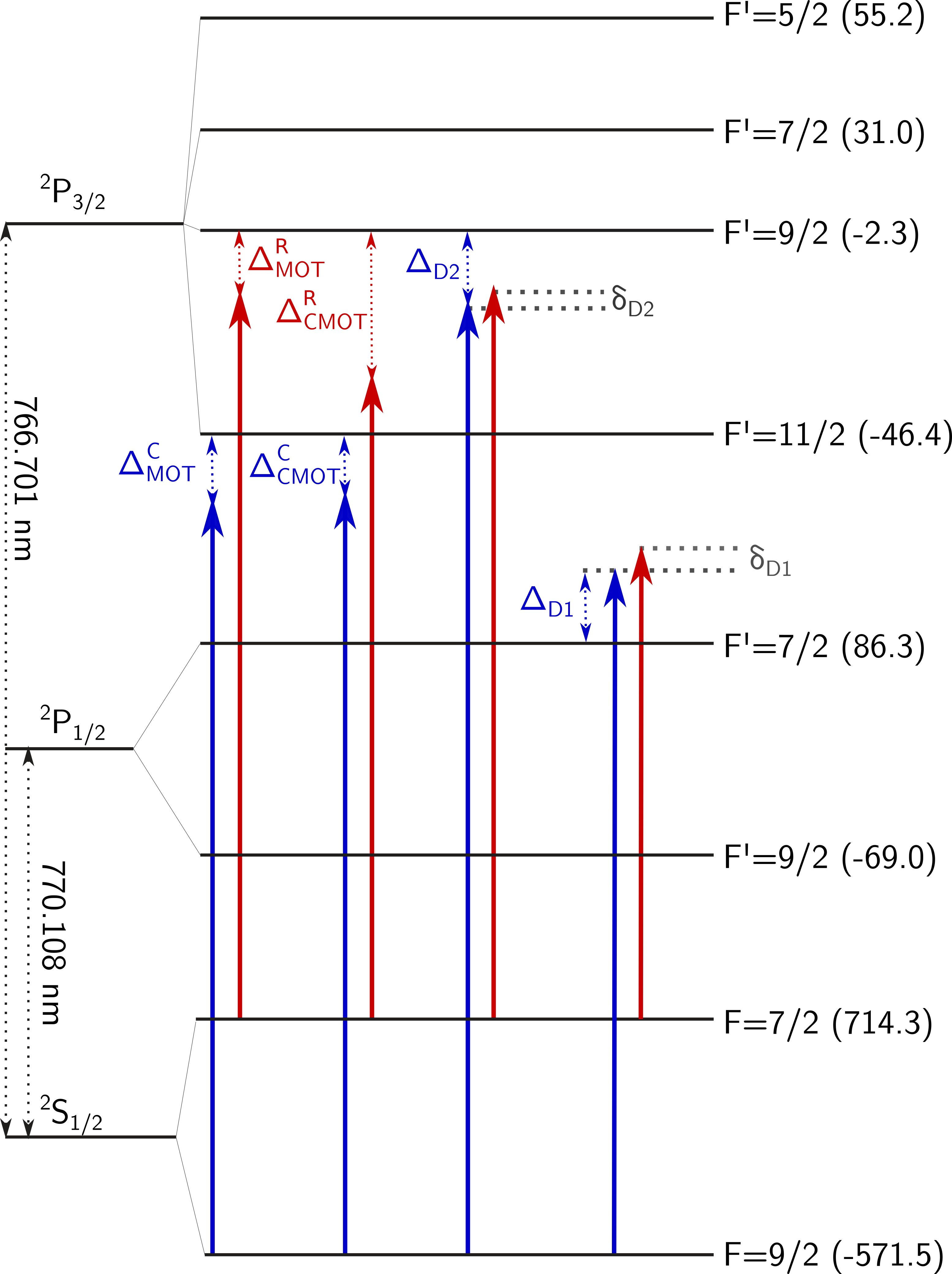}
\caption{Atomic level scheme showing the $D_1$ and the $D_2$ lines of $^{40}$K. The vertical arrows mark the beams dedicated to the cooling methods investigated in this work. For the $D_2$ line the figure shows detunings of MOT beams from the cooling transition $\Delta^\mathrm{C}_\mathrm{{MOT}}$ and from the repumping transition $\Delta^\mathrm{R}_\mathrm{{MOT}}$, detuning of the compressed MOT beams ($\Delta^\mathrm{C}_\mathrm{{CMOT}}$ and $\Delta^\mathrm{R}_\mathrm{{CMOT}}$), single-photon detuning of gray molasses beams $\mathrm{\Delta_{D2}}$ and the detuning from the two-photon resonance $\mathrm{\delta_{D2}}$. For the $D_1$ line we use, by analogy, single-photon detuning of gray molasses beams $\mathrm{\Delta_{D1}}$ and the detuning from the two-photon resonance $\mathrm{\delta_{D1}}$. The colors blue and red refer to the cooling and to the repumping beam, respectively. The hyperfine shifts are in units of MHz and are based on~\cite{18}.}
\label{fig:HFS}
\end{figure}
\section{Magneto-optical trap}

For magneto-optical trapping we use well-established laser cooling methods where we close the $^2 S_{1 / 2}, F= 9 / 2 \to {^2P_{3 / 2}}, F'=11 / 2$ cooling transition by providing repumping light on the $^2 S_{1 / 2}, F= 7 / 2 \to {^2P_{3 / 2}}, F'=9 / 2$ transition to depopulate the ground hyperfine state with lower total angular momentum $F$. It should be noted that this level has a higher energy than the $F=9/2$ level, unlike other alkali atoms. In figure~\ref{fig:HFS} we present the relevant energy levels and schematically show various detunings of the laser beams used at different stages of laser cooling on both the $D_1$ and the $D_2$ line. 

\begin{figure}[ht]
\centering
\includegraphics[width=0.9\columnwidth]{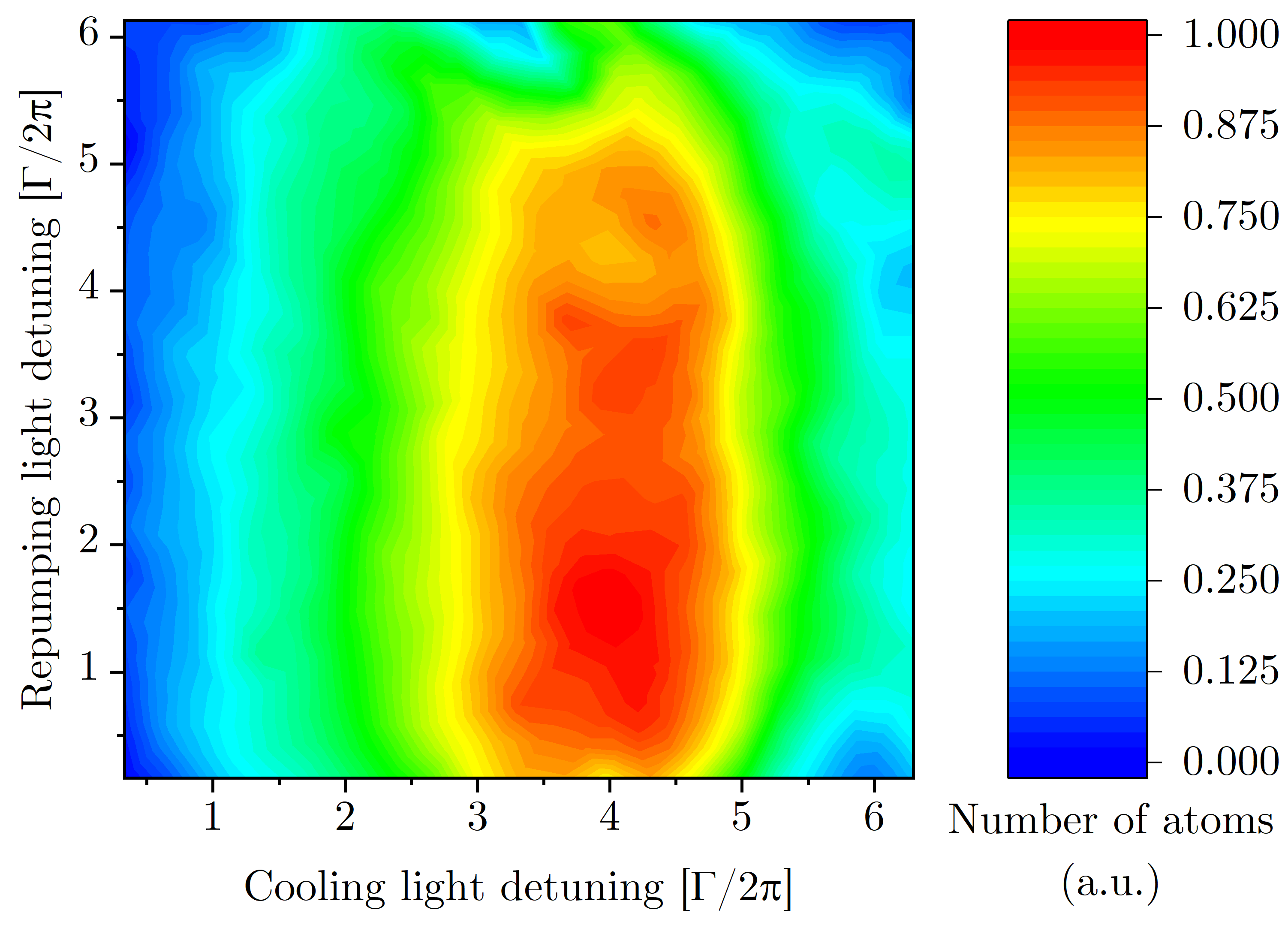}
\caption{Normalized number of atoms as a function of red-detuning from the cooling and the repumping transitions. Here, $\Gamma\approx2\pi\times6.035$~MHz is the natural linewidth of the $D_2$-line.} 
\label{fig:mot_optimal_freq}
\end{figure}
During MOT loading we use maximum available laser power with total intensities of cooling (repumping) beams equal to 29$I_\mathrm{s}$ (18$I_\mathrm{s}$), where $I_\mathrm{s}=1.75$~mW\,cm$^{-2}$ is the saturation intensity of the $D_2$ line (used also for the $D_1$ line). We have found that we can obtain the highest atom number in the MOT for the magnetic field gradient of 11~G\,cm$^{-1}$, the cooling light red-detuned by \mbox{$-3.5\Gamma$~to~$-4.5\Gamma$} and the repumper red-detuned by \mbox{$-0.5\Gamma$~to~$-4\Gamma$}~(see figure~\ref{fig:mot_optimal_freq}). Here, $\Gamma\approx6.035$~MHz is the natural linewidth of the $D_2$ line transitions. Throughout the article we also use this value for the $D_1$ line. All measurements reported in the following paragraphs are obtained for the cooler (repumper) detunings \mbox{equal to $-4\Gamma~(-1.5\Gamma$)}.

\begin{figure}[ht]
\centering
\includegraphics[width=0.95\columnwidth]{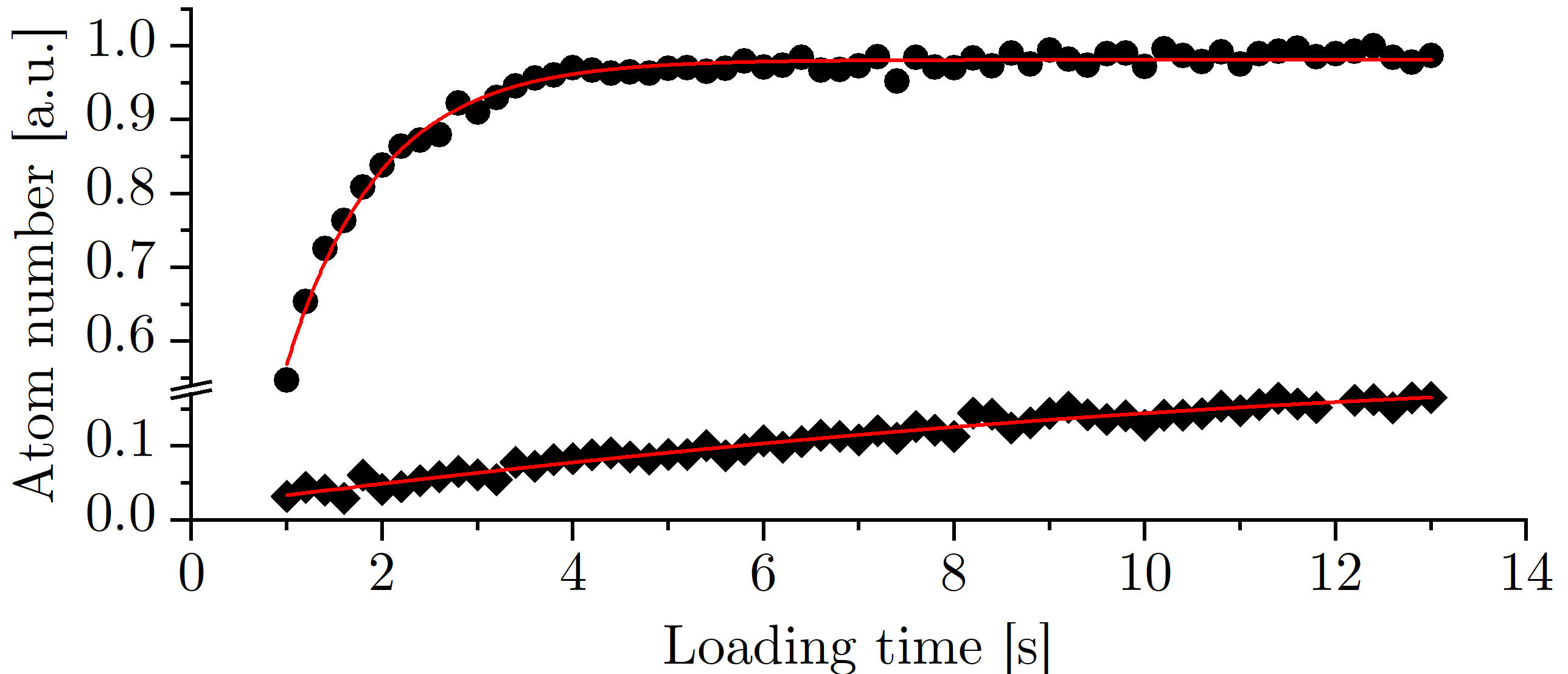}
\caption{Magneto-optical trap loading curves normalized to the maximum trappable atom number. With dispensers and LIAD \textit{turned on} (dots) the maximum number of atoms is $N_{\mathrm{max}}=5.5 \times 10^5$. With dispensers \textit{turned off} and LIAD \textit{turned on} (diamonds) this number drops to $N_{\mathrm{max}}= 8.5 \times 10^4$.} 
\label{fig:loading_curves}
\end{figure}

With optimized parameters, we have compared MOT loading curves for two cases relevant to this investigation (see figure~\ref{fig:loading_curves}). In the first case, the number of atoms is maximized by increasing the dispenser current to 4~A, thus heating it beyond typical temperatures we use while working with nearly {8,000} times more abundant $^{39}$K. The MOT loading rate is further enhanced with LIAD. Increased temperature of the dispenser reduces the quality of vacuum (see section~\ref{sec:magn_trap}) as it releases significant amounts of other, more abundant, potassium isotopes that collide with trapped $^{40}$K. Additionally, due to the nearby location of cesium dispensers, we also observe an increase in the partial pressure of cesium, enhanced even more by the use of LIAD. For regular operation, when working with $^{39}$K or Cs or their mixture, this crosstalk is minimized by choosing the temperature of the dispenser (corresponding to 3.6~A current) that provides a satisfactory compromise between the number of atoms and the lifetime of the atoms trapped in an optical dipole trap ($\sim$5~s). For the study of sub-Doppler cooling, which takes on the order of 15~ms, this reduced lifetime is irrelevant, thus the efficiency of the cooling process is investigated under these conditions.

The second case considers loading the MOT using only LIAD, with atomic sources having been turned off for more than a day resulting in a much improved background pressure. As expected, the trapped atom number is significantly smaller but the lifetime of trapped atoms is longer (see section~\ref{sec:magn_trap}). However, the overall performance of sub-Doppler cooling is essentially the same as in the first case considered.

Under conditions optimized for the maximum atom number we achieve a loading rate of $5\times10^5$~atoms\,s$^{-1}$, trapping about $5.5\times10^5$ atoms, nearly two orders of magnitude more than in the first reported MOTs of fermionic potassium~\cite{6,7}. When the dispensers have been off for at least a day and only LIAD is used, the MOT loading rate drops to $1\times10^4$~atoms\,s$^{-1}$ and we can trap at best $8.5\times10^4$ atoms. The steady-state temperature of the MOT is 250~\textmu K, well above the Doppler limit of $T_{\mathrm{D}}=145$~\textmu K~\cite{18} indicating that sub-Doppler cooling mechanisms that are typically present in MOTs of $^{40}$K~\cite{19,20,1, 7} do not seem to work in our setup. We have determined that this discrepancy is caused by the waveplates we use to set the polarization of the MOT (and gray molasses): their settings are optimized for $D_1$ line cooling which deteriorates the quality of polarization for cooling on the $D_2$ line. By fine-tuning the polarization of the MOT beams we can reach ~170~\textmu K, somewhat lower than reported for example in~\cite{19} and~\cite{10}. However, this leads to inefficient cooling on the $D_1$ line. The figure of merit for us has been the combined efficiency of all cooling stages, not of each individual stage, and, as we show next, 250~\textmu K is a sufficiently good starting point to proceed with further cooling. 

\begin{figure*}[t!]
\centering
\includegraphics[width=1.95\columnwidth]{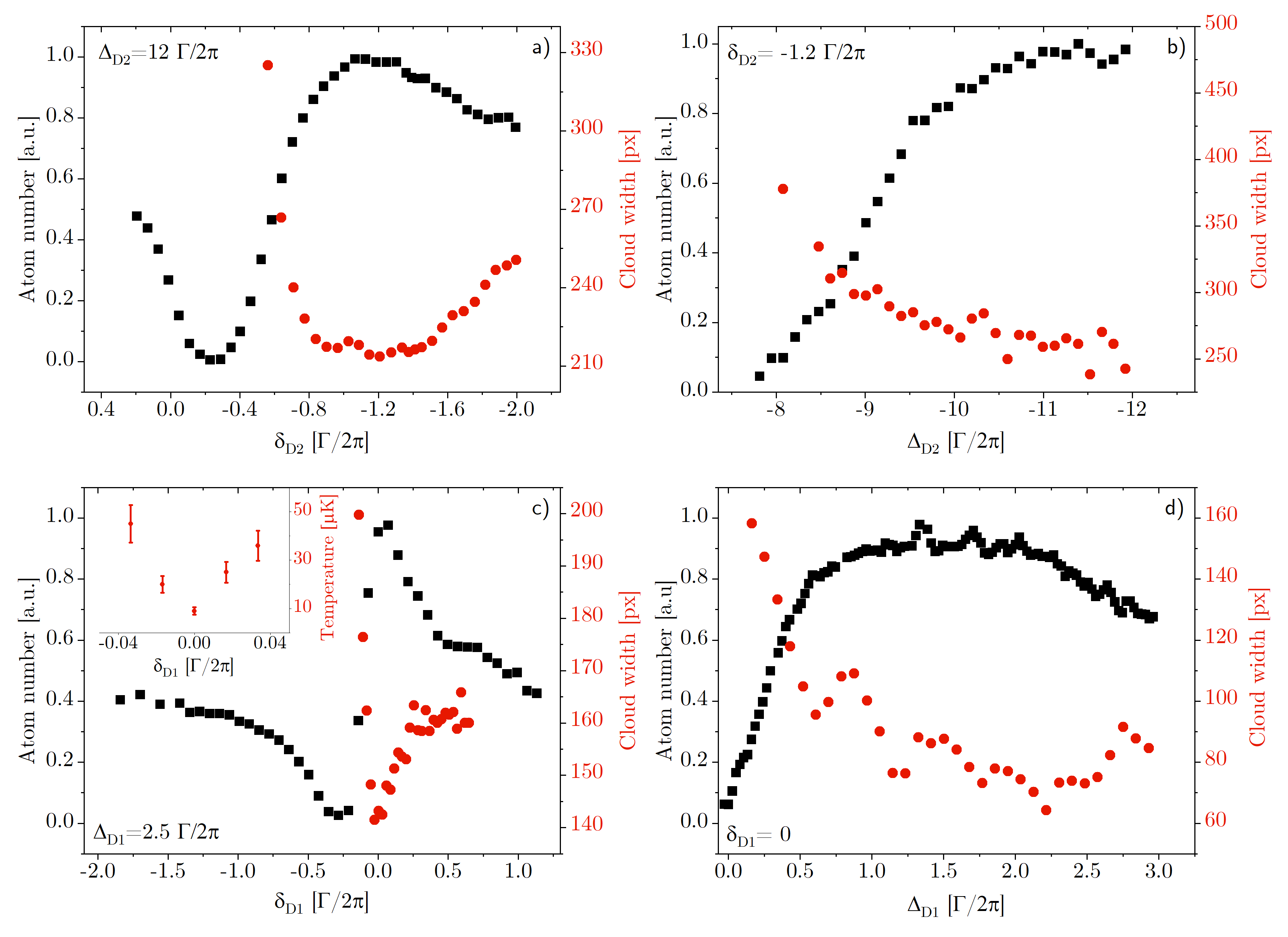}
\caption{Plots of the cloud width (red dots) and the number of atoms (black squares) as a function of (a) detuning from the Raman (two-photon) transition and (b) red-detuning (single--photon detuning) from the $^2P_{3/2}$, $F=9/2$ level obtained after 8~ms of cooling. Analogous measurements for the $D_1$ line gray-optical molasses cooling lasting 10~ms as a function of (c) detuning from the Raman (two-photon) transition and (d) blue-detuning from the $^2P_{1/2}$, $F=7/2$ level. For each plot, we show the value of the detuning that is fixed. Inset in \ref{fig:GMCD2}(c): the temperature measured for five two-photon detunings in close proximity to the two-photon transition on the $D_1$ line, obtained with optimized single-photon detuning and optimized cooling powers.} 
\label{fig:GMCD2}
\end{figure*}
\section{Sub-Doppler cooling}

We compress the cloud to cool it down and increase its density using a $D_2$-line compressed MOT stage. This is obtained by changing the cooling (repumping) beam detuning to $-1\Gamma$ ($-5\Gamma$) and increasing the magnetic field gradient to 18~G\,cm$^{-1}$ in $<$0.6~ms (set by the response time of the coil current driver). Once the magnetic gradient has reached the new value we ramp down the power of the cooling and the repumping beam in 10~ms to 6$I_\mathrm{s}$ and 1.4$I_\mathrm{s}$, respectively. We are not able to simultaneously ramp the frequency and the power due to the limitation of the RF source driving AOMs. This might be the reason that we do not observe a significant temperature drop, ending up with a cloud at 160~\textmu K. However, during this process there is no significant loss of atoms and we reach a phase space density (PSD) of $\mathrm{\rho = 1.6\times10^{-7}}$. 

We first investigate to what extent our setup allows sub-Doppler cooling on the $D_2$ line. Previous works using the $D_2$ line reached temperatures below 50~\textmu K in optical molasses with the cooling beam red-detuned by more than $-4\Gamma$~\cite{9,10} from the $F'=11/2$ level or by using coherent superposition of ground hyperfine states with red-detuned gray molasses~\cite{21}. 
We essentially follow the work of Bruce~\textit{et~al}~\cite{21} where $D_2$-line gray molasses for $^{40}$K have been implemented for the first time. After the magnetic field of the compressed MOT has been switched off, we set the compensation coil currents to nullify the background magnetic field~\cite{22} and detune the repumping and the cooling light to produce a coherent superposition of ground-state hyperfine levels via the excited state $F'=9/2$. We investigate the dependence of the final atom number and the final temperature on both the single-photon detuning from the $F'=9/2$ level, $\Delta_\mathrm{D2}$, and on the two-photon detuning $\delta_\mathrm{D2}$ between the cooling and the repumping light. Here, we have used the cloud size after a fixed expansion time as a proxy for temperature to generate the plots shown in figures~\ref{fig:GMCD2}(a)~and~(b). We find that the maximum atom number and the minimum temperature can be obtained after 8~ms of cooling for the two-photon detuning of approximately $\delta_\mathrm{D2} = -1\Gamma$ and the single-photon detuning of approximately $\Delta_\mathrm{D2}=-12\Gamma$, very similar to values reported by Bruce~\textit{et~al}~\cite{21}. Unfortunately, we observe a nearly 80\% loss of atoms with a negligible temperature drop to~140~\textmu K, as determined with a time-of-flight method. We have determined that this is primarily caused by optimization of the cooling paths for the 770~nm light, i.e. $D_1$-line light. MOT beams and gray molasses beams are coupled into the same optical fiber and share the same paths when the light leaves the fiber to form a MOT. We control the polarization of the overlapped beams with half-~and quarter-waveplates with the design wavelength of 767~nm ($D_2$ line). We optimize the polarization of the 770~nm light, which slightly degrades the target polarization of the MOT beams and $D_2$ line molasses, leading to observed inefficiencies. It is not surprising that under such conditions the sub-Doppler cooling mechanisms on the $D_2$ line do not perform as well as reported by other groups~\cite{9,10,21}. When waveplates are optimized for gray molasses of the $D_2$ line, we can cool atoms to 70(4)~\textmu K, slightly colder than 80(1)~\textmu K reported in~\cite{21}. At the same time, the cooling on the $D_1$ line becomes very inefficient: the coldest cloud reaches $\sim$80~\textmu K. Most of the sub-Doppler cooling techniques rely on simultaneous ramping of the laser frequency and power. This feature cannot be implemented with our current RF sources driving AOMs and this is the most likely reason why we could not match 48~\textmu K achieved by Bruce \textit{et~al}~\cite{21}.  The issue of the poor performance of the $D_2$ line gray molasses is inconsequential for our measurements -- when constructing the laser system, we had expected to use sub-Doppler cooling on the $D_1$ line all along while treating the MOT as a source of pre-cooled atoms. As such, the experimental setup has been optimized to maximize the atom number in the MOT and minimize the final temperature of the cloud after $D_1$-line cooling. Potential improvements could be achieved if achromatic waveplates were used.

\begin{figure}[ht]
\centering
\includegraphics[width=0.95\columnwidth]{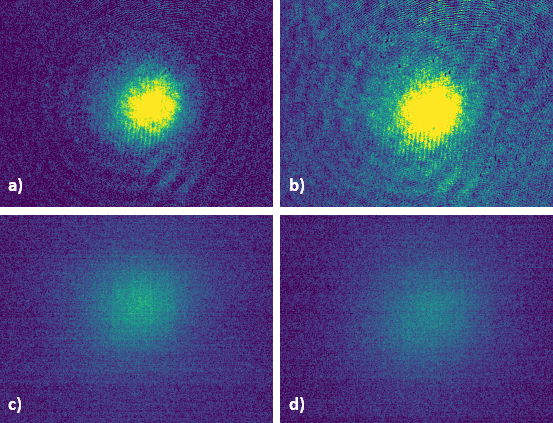}
\caption{Examples of absorption (a),(b) and fluorescence (c),(d) images taken 4.6~ms (a),(c) and 7.6~ms (b),(d) after turning off the $D_1$-line gray molasses. Due to the timing of the shutters used in the laser systems, the expansion time could not be shorter than 4.3~ms. The upper limit on expansion time is only slightly larger than 7.6~ms due to the loss of optical density resulting from the small atom number.} 
\label{fig:abs_fluo}
\end{figure}

Now we focus our attention on the gray molasses cooling on the $D_1$ line. It has been shown by several authors~\cite{15,16,23,19} that this cooling method assures highly efficient and fast cooling with negligible atom loss. In fact, it has been implemented already for all alkali atoms where sub-Doppler cooling mechanisms on the $D_2$ line do not work efficiently due to the small hyperfine structure energy splitting of the $P_{3/2}$ state. Gray molasses require good control over the background magnetic field, and it is necessary to cancel all external stray magnetic fields for the best performance. We have nullified stray fields using a method we have developed and illustrated with $^{39}$K~\cite{22} which has enabled us to cool that isotope to $\sim$8~\textmu K. Gray molasses cooling starts immediately after the $D_2$-line compressed MOT. The magnetic field and the $D_2$ light are turned off, and the compensation of the stray magnetic fields is engaged. We block the light from the master laser (D2 ML) with the shutter S2, simultaneously opening the shutter S1 that sends the $D_1$ light to the tapered amplifiers. Due to the opening and closing times of shutters, and their jitter, the $D_1$ gray molasses stage cannot start sooner than 0.8~ms after turning off the magnetic field. During that time, the cloud drops freely and becomes diluted due to its rather high temperature, but we are able to achieve a transfer of almost 100\% of atoms to gray molasses with light intensities of 8.7$I_\mathrm{s}$ and 2.2$I_\mathrm{s}$ for the cooling and the repumping beam, respectively. We proceed with optimization of single- and two-photon detuning (see figures~\ref{fig:GMCD2}(c) and(d)) and the cooling time. We have found that during the 10~ms cooling stage when we simultaneously ramp down the cooling (repumping) beam intensity from 8.7$I_\mathrm{s}$ to 2.9$I_\mathrm{s}$ (2.2$I_\mathrm{s}$ to 0.8$I_\mathrm{s}$) we are able to decrease the temperature to (10$\pm$2)~\textmu K, on par with temperatures reported by other groups~\cite{19,20} while losing less than 20\% of atoms. With nearly $3\times 10^5$ atoms in a mixture of states we obtain a free-space phase-space density $\rho = 3.5\times 10^{-5}$. For the determination of the final temperature we have used a time-of-flight method using both fluorescence and absorption images (examples of both types of images taken at different expansion times are shown in figure~\ref{fig:abs_fluo}). Both methods have shown very good agreement, giving essentially the same temperature of (10$\pm$2)~\textmu K. Here, the shortest expansion time after release from gray molasses is 4.3~ms, limited by the 14~ms repetition time of mechanical shutters S1 and S2 (see figure~\ref{fig:laser_system_simplified_K}), whereas the longest expansion is 9~ms, limited by our ability to reliably image diluted samples.

\section{Magnetic trapping\label{sec:magn_trap}}

The high laser cooling efficiency of the sample allows us to transfer atoms to a conservative magnetic potential. A magnetic trap has a large capture volume and could facilitate sympathetic cooling of $^{40}$K with $^{23}$Na, $^{41}$K or $^{87}$Rb~\cite{24,25,12}. In such a cooling scheme, the final temperature of fermions is set by the ability to evaporate the coolant (e.g. $^{23}$Na, $^{41}$K or $^{87}$Rb), and one can enter the degenerate regime without notable loss of $^{40}$K atoms.

To transfer atoms from gray molasses to the magnetic trap we perform hyperfine pumping of the atomic population to the \mbox{$^2S_{1/2}, F=7/2$} hyperfine state by switching off the repumping light, then after 500~\textmu s switching off the cooling light and turning on the quadrupole field. We use the highest gradient we can safely sustain in our setup, $B'_{\mathrm{r}}=57.6$~G\,cm$^{-1}$ (note that the horizontal gradient is nearly $B'_{\mathrm{axial}}=115$~G\,cm$^{-1}$) which is sufficiently high to capture atoms distributed between $m_\mathrm{F}=-3/2,-5/2,-7/2$ states. If after hyperfine pumping the population was equally distributed between Zeeman sublevels we would expect a 37.5\% transfer efficiency -- we obtain 16\%.

\begin{figure}[ht]
\centering
\includegraphics[width=0.95\columnwidth]{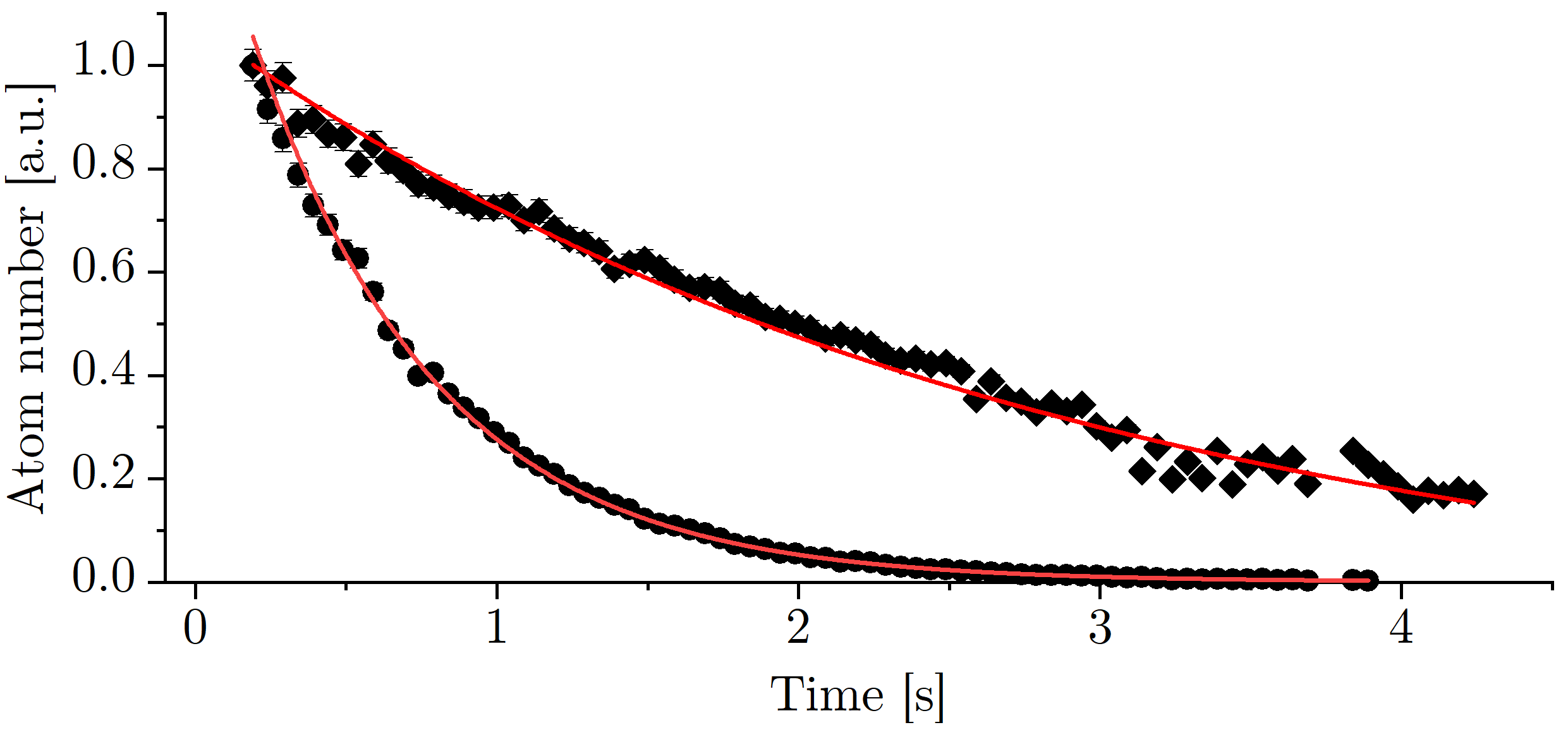}
\caption{Lifetime of $F=7/2$ potassium in a magnetic trap. Circles (diamonds) represent data normalized to the initial trapped atom number with (without) dispensers turned on. The decay is fitted with exponential functions with time constants $\tau_1=0.61$~s (circles) and $\tau_2=2.82$~s (diamonds).} 
\label{fig:lifetime}
\end{figure}

We have investigated the same two cases that we have considered for MOT loading: dispensers turned on to maximize the trapped atom number and dispensers turned off for a day and loading enhancement with LIAD. For the atom-number maximized case we are able to magnetically trap about $5\times10^4$ atoms, with a lifetime of 0.6~s. This lifetime can be extended to over 2.8~s after keeping the dispensers off for a day and enhancing the loading rate with LIAD. However, this results in a much lower number of trapped atoms, just above 3,000. The lifetime measurements are shown in figure~\ref{fig:lifetime}.
We have not implemented spin polarization as we have focused on a proof-of-principle demonstration of magnetic trapping but there is nothing fundamental nor technical that would prevent us from trapping a little over $10^5$ spin polarized atoms in our current setup under the condition of a reduced lifetime. It is worth emphasizing that the reported lifetime is a lower bound on what can be achieved in our setup as both the sample's spin polarization and/or trapping atoms in the $F=9/2$ hyperfine state would lead to the increase of the storage time. 
 
We have not been able to reliably measure the temperature of the magnetically trapped cloud as it becomes diluted after a short expansion time. This might indicate that the transfer from gray molasses introduces some heating, which would not be surprising given the nature of optical pumping and the fact that the gray molasses is slightly offset from the minimum of the magnetic potential. However, we believe that it is primarily due to the small number of atoms that approaches the detection sensitivity of our imaging system. For the lifetime measurement in the magnetic trap, the small atom number is less of an issue. We release atoms after a given hold time and recapture them in the MOT, where the fluorescence signal is collected for 20~ms. The imaging time has been chosen such that loading of the MOT from the background gas during imaging is negligible.

\section{Summary}

We have presented sub-Doppler cooling of $^{40}$K using a source with natural composition of potassium isotopes and a single chamber apparatus. With $3 \times 10^5$ atoms at a temperature of (10$\pm$2)~\textmu K after $D_1$ gray molasses cooling stage we demonstrate that even without isotopically enriched sources it is possible to achieve state-of-the-art cloud parameters in a single-chamber setup. Our results open doors to using unenriched fermionic potassium in modern experiments including quantum computing with atoms in optical tweezers and creation of ground-state molecules containing $^{40}$K if magnetically trapped atoms are sympathetically coolled with another species. For the proof-of-principle work in optical tweezers, including the formation of ground-state $^{40}$KCs molecules that we are pursuing, the vacuum quality might be a secondary issue. However, the observed deterioration of the vacuum quality when atomic sources are operational can be a serious problem in many experiments involving evaporation. As a remedy, optical or magnetic transport can be used to move the sample to another chamber with much better vacuum as has been demonstrated both for atoms~\cite{26,27} and for molecules~\cite{28}, where the cloud has been moved by 46~cm in just 50~ms.

Our work might encourage precision spectroscopic measurements of stable potassium isotopes~\cite{29} in simplified experimental setups (no enriched sources, no Zeeman slowers, no 2D MOTs etc) while still providing samples at $\sim$10~\textmu K thus significantly reducing the Doppler effect. Our results on sub-Doppler cooling of small numbers of atoms might be useful to $\beta$-decay experiments, where similarly sized samples of laser-cooled $^{37}$K and $^{38m}$K isotopes have been used~\cite{30}. Here, it is not only a matter of the reduction of the Doppler effect but, we believe, of the demonstrated gain in phase-space density that might improve the quality of measurements. 

In the same setup as used in this work we routinely cool 5$\times10^7$ $^{41}$K to $\sim$10~\textmu K, achieving lifetimes of magnetically trapped samples of several seconds. The content of $^{41}$K in natural potassium is~6.7\%, therefore if we were to use a source with $^{40}$K enriched to 5\% (e.g. dispensers from AlfaVakuo e.U.) we could expect more than $10^7$ fermionic atoms at $\sim$10~\textmu K after gray molasses. After implementing spin polarization we could reasonably expect to magnetically trap $>10^6$ $^{40}$K atoms with a lifetime of $>5$~s while still using a very simple single-chamber design.

In a related work performed using the same experimental setup we have achieved nearly 90\% transfer efficiency of $^{40}$K from the $D_1$-line gray molasses to the magnetic trap operating at 57.6~G\,cm$^{-1}$. The transferred sample has been spin-polarized in the $F = 9/2,\,m_\mathrm{F} = 9/2$ state, with the lifetime reaching 1.5(1)~s.\textbf{}

%\acknowledgments 
We would like to acknowledge P Arciszewski and J Dobosz for their contribution to the development of the experimental setup used in this work. This research was funded by the Foundation for Polish Science within the HOMING programme and the National Science Centre of Poland (Grant No. 2016/21/D/ST2/02003 and a postdoctoral fellowship for M~S, Grant No. DEC-2015/16/S/ST2/00425).
%\typeout{}
%\printbibliography

\end{document}